\documentclass[12pt]{article}
\usepackage[latin9]{inputenc}
\usepackage{amstext}
\usepackage[unicode=true]
 {hyperref}

\makeatletter

\newcommand{\pubnumber}{RM3-TH/19-1}
\newcommand{\pubdate}{\today}

\def\institute{Universit\`a degli Studi Roma Tre and INFN Roma Tre\\
Via della Vasca Navale, 84 I-00146 Roma, ITALY}
\def\support{\footnote{ This work was supported by Programma per Giovani Ricercatori Rita Levi Montalcini granted by Ministero dell'Istruzione, dell'Universit\`a e della Ricerca (MIUR).}}

\def\Title#1{\begin{center} {\Large #1 } \end{center}}
\def\Author#1{\begin{center}{ \sc #1} \end{center}}
\def\Address#1{\begin{center}{ \it #1} \end{center}}

\newcommand{\pubblock}{\rightline{\begin{tabular}{l} \pubnumber\\
         \pubdate  \end{tabular}}}
\newenvironment{Abstract}{\begin{quotation}  }{\end{quotation}}
\newenvironment{Presented}{\begin{quotation} \begin{center} 
             PRESENTED AT\end{center}\bigskip 
      \begin{center}\begin{large}}{\end{large}\end{center} \end{quotation}}

\def\beq{\begin{equation}}
\def\eeq#1{\label{#1}\end{equation}}
\def\eeqn{\end{equation}}

\def\beqa{\begin{eqnarray}}
\def\eeqa#1{\label{#1}\end{eqnarray}}
\def\eeqan{\end{eqnarray}}

\let\bar=\overbar

\def\Dslash{\not{\hbox{\kern-4pt $D$}}}
\def\dslash{\not{\hbox{\kern-2pt $\del$}}}

\def\msb{{\bar{\ssstyle M \kern -1pt S}}}

\makeatother

\begin{document}
\begin{titlepage} \pubblock

\vfill{}
\Title{The Higgs Top Interface} \vfill{}
\Author{ Roberto Franceschini\support} \Address{\institute}
\vfill{}
\begin{Abstract} After some recollection of the implications that
top quark and Higgs boson properties determination have had on each
other, we discuss recent and future expected progress in the study
of the top-Higgs sector. In particular we discuss some results concerning
new physics states interacting with the top-Higgs sector, the impact
of measuring top-Higgs interactions accurately to determine the Higgs
boson self interaction, the CP nature of the Higgs boson and possible
detection of flavor violating interaction at levels comparable with
the prediction of viable new physics models. \end{Abstract} \vfill{}
\begin{Presented} $11^{\mathrm{th}}$ International Workshop on Top
Quark Physics\\
 Bad Neuenahr, Germany, September 16--21, 2018 \end{Presented} \vfill{}
\end{titlepage} \global\long\def\thefootnote{\fnsymbol{footnote}}
 \setcounter{footnote}{0} 

\section{Introduction}

Top quarks and Higgs boson interact in the Standard Model via Yukawa
interactions; this is, literally speaking, the top-Higgs interface.
Its nature is still largely unknown, though current experiments have
started to probe it and future ones promise to do so with a finer
level of detail.

Basic issues about this interaction have to do in part with fingerprinting
the objects that participate to it, and in part with the understanding
of the origin of this interaction \emph{per se}. Remarkably, at the
top-Higgs interface it is possible to probe the nature of the Higgs
boson in a new way, independent of the many tests performed so far
in Higgs boson studies carried out at the LHC. Properties of the Higgs
boson such as its CP-nature may be probed in new ways. The gauge charge
of the fields that gives rise the physical Higgs boson can be probed
as well. Furthermore it is possible to explore the nature of the top
quark, which may be the first quark to show signs of compositeness,
that is to say to not be a point-like field, and its size can be established
through the magnifying glass of the Higgs boson probe. 

In addition, the UV origin of the top-Higgs Yukawa interaction needs
to be established in order to claim some satisfactory degree of understanding
of the Standard Model as a whole. A key question on the nature of
the top-Higgs Yukawa interaction has to do with it being a marginal
coupling, or having a steeper RG-flow at higher energy scales. For
what we know this interaction may originate from a contact interaction
involving more than one scalar, possibly related to the dynamics of
flavor, or even be a low energy artifact from a more complete theory
in which the Yukawa is a relevant interaction only at the long length-scales
we currently probe in our still too-low energy experiments.

In addition to this ``physical'' connection, the Higgs boson and
the top quark have an intertwined history, due to the strong theoretical
relation that arises in the Standard Model between these two particles.
For instance, the effect of the mass of one particle in radiative
corrections to the mass of the other made possible to predict, in
the pure Standard Model, where the Higgs boson mass would be within
about 10\% accuracy. Similarly, today, with the Higgs boson mass measured
at few parts per one thousand, the top quark mass can be extracted
from the rest of the parameters of the Standard Model to an accuracy
that competes with the that of best direct measurements of today.
The pretty consistent picture that emerges from this test of the Standard
Model prediction is already putting bounds on generic new physics
entering in radiative corrections and therefore affecting indirectly
some measured quantities. 

On top of this probe of generic new physics, the top-Higgs interface
motivated deep questions on the validity of the Standard Model as
ultimate theory of Nature, valid up to the shortest length-scale.
On a very concrete ground the Higgs-top interaction may modify the
Higgs boson potential to such a large extent that the potential may
develop new minima at field values much larger than the ones at which
it has condensed in the Standard Model, implying a drastic change
for our picture of fundamental interaction and possibly changing the
history of the Universe, both in its early phase, where these large
field values may be easily attained, or in the future, when our ground-state,
if meta-stable, may tunnel to the true energy minimum of the theory,
erasing the Universe as we know it. 

On a possibly more philosophical level the Higgs-top interaction has
raised questions about the possible sensitivity of weak interactions
to details of the UV theory valid at the shortest-distances. The structure
of the Standard Model seems to open the door to a possible hyper-sensitivity
to microscopic changes in the short-distance degrees of freedom. Such
a sensitivity would defy the common sense of decoupling of length-scales
that lies at the foundation of the effective field theory paradigm
on which we rely to apply quantum field theories to the natural world.
The exploration of the top-Higgs interface, and of the extensions
of the Standard Model that it has motivated, is a key tool to understand
if decoupling of length-scales applies for weak interactions at the
shortest distances we explore in high-energy physics.

\section{Recent and future highlights}

The discovery of the Higgs boson has provided already a number of
tests of the structure of the Standard Model and constraints of its
possible extensions. Precise measurements of the mass as well as the
rates in each decay channel of the Higgs boson provide stringent bounds
on new physics, or, looking from a different angle to these measurements,
can provide hints on where to find new physics if it is out there. 

\subsection{Probes from the Higgs boson mass and inclusive rates}

Extensions of the Standard Model that prevent TeV-scale physics to
depend strongly on the dynamics of far higher energy scales usually
have the property of being able to compute the Higgs mass as a function
of few parameters of the theory. These parameters include those of
the Standard Model and few extra ones, which also affect other measurable
quantities. An example is the case of the MSSM in which the Higgs
boson mass can be computed once the properties (masses and mixings)
of the scalar top and bottom partners are known. As shown in Ref.~\cite{Dermisek:2008rt},
the values of the Higgs boson mass considered in the time between
the validation of the SM at LEP and the discovery of the Higgs boson
were significantly smaller than what observed. Plots from Ref.~\cite{Dermisek:2008rt}
indeed fall one GeV shorter than the measured value, but can be easily
extrapolated (or one can look at post-discovery references such as
Ref.~\cite{Hall:2011rt}). In these plots one can see how the measured
value of the Higgs boson mass, 125 GeV, cannot be accommodated in
the MSSM unless the scalar top quarks are above 1 TeV, no matter what
other MSSM parameters we pick. In view of this observation it appears
not surprising at all that we did not find any sign of new states
so far at the LHC. The mass range hinted by the value of the Higgs
boson mass for where to find new scalar top quarks is just now being
probed, with first exclusions of scalar top and bottom quarks masses
extending above 1 TeV. It is remarkable that, at least under the assumption
of a Higgs-top interface à la MSSM, the mere knowledge of the Higgs
boson mass has preempted direct searches for scalar top quarks. Equally
interesting is the fact that for the Higgs mass to match the observed
one the mixing in the scalar top sector has to be picked in a somewhat
special value. This defines some kind of ``sweet spot'' for the
MSSM after the discovery of the Higgs boson and the measurement of
its mass that has the lightest possible physical scalar top quark
masses $m_{\tilde{t}_{1}}\sim m_{\tilde{t}_{2}}\simeq1.5\text{ TeV}$
and the mixing angle being large, $X_{t}/m_{\tilde{t}}\simeq\pm\sqrt{6}$
in the notation of Refs.~\cite{Dermisek:2008rt,Hall:2011rt} and
references therein. It is remarkable that this choice of large mixing
is also approximatively the choice that makes any virtual effect of
scalar tops almost disappear from the production cross-section of
the Higgs boson in gluon fusion process at the LHC. Indeed computing
the combined effect on gluon-Higgs and photon-Higgs boson couplings
one can see that in the MSSM for $m_{h}\simeq125\text{ GeV}$ the
total rate $gg\to h\to\gamma\gamma$ varies at most by around 5\%
with respect to the SM value~\cite{Hall:2011rt}. This level of deviation
from the SM in the Higgs couplings measurements is in the domain of
LHC sensitivity~\cite{Cepeda:2650162} and is therefore a valuable
observable to test the Higgs boson, the top quark and the potential
existence of a larger sector of states that interact primarily with
them. Future collider facilities under discussion in these days have
all a rich Higgs boson program and promise to have an even finer control
on the property of the Higgs boson, which may reveal, or put strong
constraints on, new physics in the top-Higgs sector.

\subsection{Direct access to the top quark Yukawa coupling and implications}

Direct observations of the $pp\to t\bar{t}h$ reaction has given a
direct measurement of the top quark Yukawa at the LHC. This opens
up a new way of exploring new physics. In fact with this measurement
it is possible to put the computation of Higgs main production $gg\to h$
on a firmer ground. The gluon-Higgs boson coupling depends linearly
on the top Yukawa couplings. This implies that a direct knowledge
of the latter allows to measure all independent combinations of couplings
of the SM in the Higgs-top sector - hence giving a chance to constrain
contributions to the $gg\to h$ process from physics beyond the Standard
Model. A similar disentanglement of effects takes place in the photon-Higgs
boson coupling. Putting these effects together allows to constrain
generic new physics particles affecting the Higgs boson loop level
couplings. In Ref.~\cite{Essig:2017aa} these bounds have been computed,
devoting also some attention to the cases in which they can be avoided
by judicious choices of other parameters, such as the mixing between
pairs of BSM states contributing to the Higgs boson loop couplings.
As we have seen for the case of the MSSM, dismissing choices of parameters
as ``ad-hoc hence uninteresting'', may be too quick a judgement
as they may be motivated by other considerations and convenient also
to explain other aspects of the Higgs boson physics. However, if one
takes the route to ignore these peculiar regions of parameter space,
it is possible to draw conclusions and exclude the existence of new
particles in the Higgs-top sector. These bounds rely crucially on
the knowledge of the top quark Yukawa, hence they demand the best
possible accuracy in the measurement of the $tth$ cross-section at
the LHC and motivate further efforts in future collides to pin down
the Yukawa coupling of the top quark as precisely as possible. By
the end of the HL-LHC we expect that these bounds may be sensitive
to scalar stops lighter than 500 GeV, independently of how it decays,
for a top quark Yukawa measured within around 10\%, fully in the reach
of HL-LHC~\cite{Cepeda:2650162}. 

The top quark Yukawa is one of the several couplings of the Standard
Model that affects double Higgs boson production. The dependence of
the rate on the top quark Yukawa and relevant effective couplings
of an effective theory for the SM up to dimension-6 operators is reported
in Refs.~\cite{Vita:2017qd,Azatov:2015oxa}. The joint study of double
Higgs production, top-Higgs boson associated production $t\bar{t}h$,
and single Higgs boson production rates allows to constrain simultaneously
three effective couplings and in particular those that result in modified
Higgs boson self-interaction. The impact of the error on the effective
coupling entering the $t\bar{t}h$ process has been studied in Ref.~\cite{Vita:2017qd},
where it is shown that Higgs self-coupling constraints can improve
by a factor around 2 thanks to the improvements on the top-Higgs interactions
coming from direct observation of the $t\bar{t}h$ process. 

The nature of the top-Higgs coupling is also a very sensitive probe
of CP properties of the Higgs boson. While these are tested already
in other processes such as Higgs boson decays to vector bosons, it
should be stressed that in concrete models the effects of CP-mixing
may be enhanced in fermionic interactions and in particular there
might be effects that depend on the flavor of the fermion. A concrete
example of this type of scenario is the model for baryogenesis discussed
in Ref.~\cite{Espinosa:2012aa}. Future HL-LHC sensitivity to CP-odd
components of the Higgs boson is discussed in Ref.~\cite{Goncalves:2018aa},
which shows how kinematical variables of the $t\bar{t}h$ final state
can be exploited to attain bounds on mixing angles larger than $\cos\alpha\simeq0.5$,
which are interesting for the generation of the baryon number of the
Universe in models such as that discussed in Ref.~\cite{Espinosa:2012aa}. 

\subsection{New processes with top quarks and Higgs bosons}

It is worth recalling new processes that may involve the top quark
and an extended Higgs sector, such as those predicted by non-minimal
models of composite Higgs scenarios. One interesting example is the
possible decay of the top quark in light flavor quarks, e.g. a charm
quark, and new Higgs bosons, $t\to\phi c$. This process has been
studied in Ref.~\cite{Banerjee:2018ab} which finds that HL-LHC can
be sensitive to branching fractions of order $10^{-4}$. Remarkably
this level of branching ratio is compatible with the expectation in
concrete models, thereby offering a chance to test flavor violation
in top quark decay, which instead turns out to be very challenging
for $h$, $Z$, and $\gamma$ final states of top decays. 

Also non-standard final states such as $pp\to th+X$ offer interesting
possibilities to study more closely the top-Higgs interface. For instance
the $thW$ and $thq$ rates are sensitive to interference effects
among SM amplitudes that entails some sensitivity to the sign of the
top quark Yukawa coupling relative to the Higgs-Vector bosons couplings~\cite{Farina:2012xp,Biswas:2013aa,Demartin:2017aa}
and has lead to conclusive exclusion of the hypothesis $y_{t}=-y_{t,SM}$\cite{CMS-Collaboration:2018ad,CMS-PAS-HIG-17-005}. 

\section{Conclusions}

The top quark and the Higgs boson are two pillars of the Standard
Model and have only been studied so far at a very coarse level. These
studies are nonetheless already yielding interesting conclusions on
the validity of the SM and posing constraints on the existence of
new physics. In some scenarios these results, e.g. the mass measurement
and the rate measurements of the Higgs boson, especially when combined,
suggest that new physics may be sitting at the TeV scale though has
properties that make it appear elusive in Higgs physics, but not necessarily
so in direct searches.

Other examples of important lessons that will be made sharper by sharpening
our knowledge of the top-Higgs coupling have been presented and include
the possible constraints on Higgs self-interactions, the CP nature
of the Higgs boson, possible links between the top-Higgs sector and
flavor violation. 

From this vast range of possible BSM scenarios to which the top-Higgs
sector is sensitive, it seems natural to argue that, whatever the
nature of the new physics, any improvement on the knowledge of Higgs
and top properties will teach us a lot on the SM and its possible
extensions. Indeed in many respects the top-Higgs sector is the cornerstone
of the SM, but its study is only at the beginning and we need to get
a more fine picture of these two particles and their interactions.
Therefore it appears very well motivated to pursue a strong program
towards the realization of future colliders that can study the top
quark and the Higgs boson producing and detecting them in copious
number.

\end{document}